\documentclass[aps, pre, subeqns.floatfix,twocolumn,amsmath,showpacs]{revtex4}
\usepackage{graphicx,epsfig}
\usepackage{subfigure}
\usepackage{hyperref}
\begin{document}
 \title{Diverse routes of transition from amplitude to oscillation death in coupled oscillators under additional repulsive link}
 \author{C. R. Hens$^1$,  Pinaki Pal$^2$, Sourav K. Bhowmick$^{1,3}$, Prodyot K. Roy$^4$, Abhijit Sen$^5$, Syamal K. Dana$^1$}
 \affiliation{$^1$CSIR-Indian Institute of Chemical Biology, Kolkata 700032, India}
% University of Oldenburg, Oldenburg, Germany}
\affiliation{$^2$Department of Mathematics, National Institute of Technology, Durgapur 713209, India}
\affiliation{$^3$Department of Electronics, Asutosh College, Kolkata 700026, India}
\affiliation{$^4$Department of Physics, Presidency University, Kolkata 700073, India}
 \affiliation{$^5$Institute for Plasma Research, Gandhinagar 382428, India}

 \date{\today}
 \begin{abstract}
 We report the existence of diverse routes of transition from amplitude death (AD) to oscillation death (OD) in three different diffusively coupled systems, those are perturbed by a symmetry breaking repulsive coupling link. For limit cycle systems the transition is through a pitchfork bifurcation (PB) as has been noted before, but in chaotic systems it can be through a saddle-node or a transcritical bifurcation depending on the nature of the underlying dynamics of the individual systems. The diversity of the routes and their dependence on the complex dynamics of the coupled systems not only broadens our understanding of this important phenomenon but can lead to potentially new practical applications.

%Evidence of diverse routes of transition from amplitude to oscillation death is found in two diffusively coupled oscillators when an asymmetry is introduced in the form of a repulsive coupling %interaction. In a synchronized set of two identical oscillators under diffusive coupling, when a repulsive coupling is added, amplitude death emerges above a critical coupling and, further a %transition to oscillation death occurs for increasing coupling strength. Using this repulsive coupling asymmetry in different types of synchronized oscillators, limit cycle as well as chaotic, we are %able to identify three different routes of transition: pitchfork bifurcation in limit cycle systems, transcritical bifurcation and saddle-node bifurcation in chaotic systems.  We report this transition %for the first time in any chaotic system. Numerical examples of the paradigmatic van der Pol limit cycle system and two different Sprott systems are presented. 

 \end{abstract}

\maketitle
%\section*{I. Introduction}

%{\bf(We should not think about the smooth transition about AD to OD rather we should think how to induce HSS in coupled system).
\par The cessation of oscillations in a system of coupled oscillators is a natural occurrence that has been known for long \cite{rayleigh} and continues to be the object of many current studies  \cite{saxena,koseska-kurths}. It can happen under a variety of circumstances such as a parameter mismatch \cite{mirolo} between the systems, a time delay in the coupling \cite{sen} between identical systems and their interplay with the strength of interaction of the systems or other innovative forms of coupling between them \cite{konishi,karnatak,resmi}.   The quenching of oscillations can be of two distinct types, namely, amplitude death (AD) \cite {sen, mirolo, kopell} and oscillation death (OD)\cite {bar-eli, ermentrout}.  In AD the system collapses to the origin 
which constitutes a stable fixed point of the system. In OD, on the other hand, the individual systems stabilize to different steady states such that the coupled system shows no oscillations. These two quenching types have also been classified \cite{koseska2, koseska-kurths, hens} as a stable homogeneous steady state (HSS) to represent the AD state and a stable inhomogeneous steady state (IHSS) to describe the OD state. In a HSS state, the coupled oscillators are stabilized to one unique equilibrium point
%either the origin or  any other equilibrium point. 
% or any other equilibrium point. This equilibrium may be the original fixed point of the uncoupled systems or a newly created one. All the coupled systems populate one stable steady state. 
whereas in the case of IHSS, the individual systems populate separate stable steady states. The study of these quenched states has important practical consequences. The HSS for example is a desirable goal for control of instabilities in various physical systems such as lasers \cite{wei}, and its robustness is an important criterion for a stable ground state in a healthy cell signalling network \cite{kondor}.  The IHSS has found useful applications in biological systems e.g. in representing cell differentiation \cite{koseska3, suzuki}, diversity of stable states in coupled genetic oscillators \cite{evgeni}, survival of species \cite{amritkar}, etc. 
\par An interesting and fundamental issue, that was recently addressed in \cite{koseska-kurths, koseska2}, concerns a transition from AD to OD in a coupled system.  In a limit cycle system consisting of two coupled  Landau-Stuart oscillators,  it was shown \cite{koseska2} that the transition occurs via a pitchfork bifurcation (PB) when a symmetry breaking perturbation in the form of a parameter mismatch is applied.  The phenomenon is very analogous to a Turing bifurcation \cite{turing, prigogine} in a spatially distributed medium.
% when an inhomogeneous perturbation in the form of a parameter mismatch is applied to two diffusively coupled Landau-Stuart systems, a transition from AD to OD  occurs via 
%the symmetry-breaking PB for varying coupling strength: a transition from a unique stable steady state to diverse stable steady states occurs.  
The PB transition has subsequently been found for other forms of coupling \cite{zou} as well, such as, time delayed coupling, dynamic coupling and conjugate coupling and has thereby widened  and generalized the findings of \cite{koseska2}. 
However it may be noted  that the above studies have all been restricted to the Landau-Stuart limit cycle system and it is by no means clear or established that this particular form of transition is generic to all systems and will hold for systems that have more complex dynamics. This is still an open question. In particular, to the best of our knowledge, there has been no investigation on the nature of the AD to OD transition in chaotic systems.  In this Letter we report on such a study where by using a symmetry breaking repulsive link we have compared the AD to OD transition in three different diffusively coupled systems -  a set of two coupled Van Der Pol (VDP) oscillators representing a limit cycle system and two sets of coupled Sprott systems representing chaotic oscillators. An important finding of our study is that while the transition in the set of coupled VDPs takes place via the PB route, the two chaotic systems show two distinct and diverse routes of transition depending on the nature of the chaotic attractor characterizing the system. For a Sprott system with a Lorenz-like butterfly attractor the transition follows a saddle-node bifurcation (SNB) route whereas for another Sprott system with a R\"{o}ssler like attractor \cite{sprott} the transition is induced by a transcritical bifurcation (TB). Thus the nature of the AD to OD transition is of a richer nature than hitherto thought of and takes diverse bifurcation routes that are dependent on the complex dynamics of the system. Our finding can help further deepening our understanding of this important phenomenon and may also have important implications for practical applications.

%distinct models an additional symmetry breaking repulsive coupling link to a set of diffusively coupled identical set of oscillators using the similar strategy of playing with the coupling strength %and a parameter mismatch. However, these studies are confined to Landau-Stuart limit cycle system only. It is important to explore if such a transition occurs in other systems especially chaotic %systems and what is the mechanism of transition if it occurs? Is it always the PB that induce such a transition?

\par We consider two identical $m$-dimensional systems, $\dot{\bf{X}}=F(\bf{X})$ and $\dot{\bf{Y}}=F(\bf{Y})$; where $\bf {X}, \bf{Y}$ $\in $ $R^m$, $F:R^m\rightarrow R^m$. Two such diffusively coupled systems with repulsive links can then be represented as,
\begin{subequations}
\begin{eqnarray}
\dot{\bf{X}}&=&F({\bf{X})}+ p_1 D({\bf {Y-X}})+q_1 Q({\bf{X+Y}})\nonumber\\
\dot{\bf{Y}}&=&F({\bf{Y}})+ p_2 D({\bf{X-Y}})+q_2 Q({\bf{X+Y}})\nonumber
\end{eqnarray}
\end{subequations}
where $p_1, p_2$, are either 0 or 1 denoting presence or absence of (unidirectional or bidirectional) diffusive coupling. The existence of additional mean-field repulsive link depends upon $q_1$ and $q_2$ which may exist symmetrically (bidirectional) if both of them have -1 values, otherwise asymmetric (unidirectional) if either of them is zero.
 D and Q are $m\times m$ coupling matrices.
 The additional repulsive link which plays the role of a symmetry breaking perturbation has been used before to study AD to OD transitions in the Landau-Stuart system \cite{hens}. Such a link  can model a local fault that may evolve in time between two neighboring dynamical systems or nodes in a network of synchronized oscillators and act as a negative feedback and thereby induce a damping effect on the neighboring oscillators \cite{dbs, hens, stefanovoska}. 
%of two neighboring dynamical nodes of the network and assumed to directly affect one of the immediate neighboring nodes with a negative feedback or a repulsive link.
%Now as examples, we choose three models, a 2D van der Pol system, a 3D chaotic Sprott system with a R\"ossler like coherent attractor %\cite{rossler} 
%and another 3D Sprott system \cite{sprott} which shows a butterfly type chaotic attractor. 

%We use numerical approach to identify the transitions in different systems since it is analytically difficult to identify even for the limit cycle van der Pol system. % although in our earlier work \cite{hens}, we could analytically determine the critical coupling of AD and OD and their transition in coupled Landau-Stuart system by coincidence. 

As our first example, we bidirectionally couple two VDP oscillators with one unidirectional repulsive link,
%\begin{subequations}
%\begin{eqnarray}
$\dot{x_1}=x_2-\epsilon_2(y_1+x_1)$,
$\dot{x_2}=b(1-{x_1}^2)x_2-x_1+\epsilon_1(y_2-x_2)$,
$\dot{y_1}=y_2$,
$\dot{y_2}=b(1-{y_1}^2)y_2-y_1+\epsilon_1(x_2-y_2)$,
%\end{eqnarray}
%\end{subequations}
where we assume $p_1=p_2=1$, $q_1=-1$, $q_2=0$ in the general coupled equation, system parameter, $b=0.5$ and $\epsilon_1$ is the strength of diffusive coupling and $\epsilon_2$ is the strength of repulsive link or negative feedback,
 \begin{eqnarray}
D=
\begin{pmatrix}
 0 & 0  \\
0 & \epsilon_1
  \end{pmatrix}
% \end{eqnarray}
%\begin{eqnarray}
;
Q=
\begin{pmatrix}
 \epsilon_2 & 0  \nonumber\\
0 & 0
  \end{pmatrix}
 \end{eqnarray}

%To get the eigen value as function of $\epsilon_1$ and $\epsilon_2$:
%\begin{eqnarray}
%\begin{vmatrix}.
%1-\epsilon_2-\lambda & 0  & -\epsilon_2  & 0 \\\
%-1 &  b-\epsilon_1-\lambda  & 0 & \epsilon_1 \\
%1 & 0 & -\lambda & 0 \\
%0 &\epsilon_1  & -1  & b-\epsilon_1-\lambda
%\end{vmatrix}
%=0
 %\end{eqnarray}

Two systems mutually interact via $x_2-, y_2-$ variables for synchronization while the repulsive link is added to the $x_1$-variable of the first oscillator only. The repulsive link thus affects one of the oscillators for the choice of $q_1=-1$ and $q_2=0$.  After coupling the trivial fixed point origin remains and in addition, two new fixed points are created, 
%%\begin{eqnarray}
%\label{vdp_bi}
\
$x_1^* =\pm\sqrt{1-\frac{1}{b\epsilon_2}}$; %\nonumber\\
$ x_2^*=\frac{\epsilon_2 x_1}{1-\epsilon_1 \epsilon_2}$;%\nonumber\\
$y_1^*=\epsilon_1 x_2^*$;
$y_2^*=0.$
%\end{eqnarray} \\
%[\it try to find out the condition for AD]
\begin{figure}[h]
\includegraphics[height=7cm,width=8.5cm]{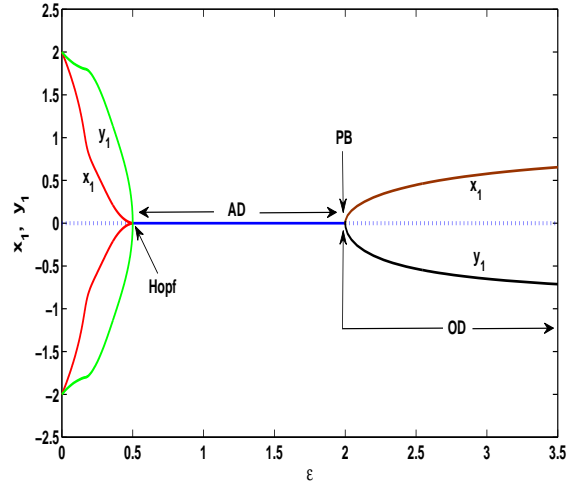}
\caption{Coupled Van Der Pol system. Extrema of $x_1$ and $y_1$ are plotted as a function of $\epsilon$; bifurcation diagram shows a transition from oscillatory state to AD via Hopf bifurcation and then a transition to OD via PB.} \label{fig:vdp_2osc_bi}
\end{figure}
For simplicity we consider $\epsilon_1=\epsilon_2=\epsilon$ and carry out a detailed numerical analysis of the system.  It is confirmed (using MatCont software package \cite{dhooge}) that AD appears at origin when $\epsilon>b$ and a transition to OD occurs via PB at $\epsilon=\frac{1}{b}$ as shown in Fig. {\ref{fig:vdp_2osc_bi}} when new equilibrium points originate for $\epsilon>\frac{1}{b}$. 
%{\it Now we  calculate largest real part of eigen values and draw a phase space plot as a function of $\epsilon_1$ and $\epsilon_2$.}
AD appears at left via reverse Hopf bifurcation (HB) where the extrema of limit cycle (green and red lines represent 
$x_1$ and $y_1$ variables) reach the horizontal zero line at stable origin and this continues (blue line) until it breaks into
 two solid lines (black and brown) at the PB point and, the stable origin (dashed blue line) becomes unstable. Beyond the PB point, the coupled systems populate two newly created stable fixed points. 
\par  We next consider a chaotic system instead of a limit cycle system. The basic uncoupled system is a Sprott system whose model is given by $\dot{x_1}=x_2 x_3$; $\dot{x_2}=x_1-b x_2$; $\dot{x_3}=a-x_1 x_2$ ($a=1$, $b=1$ produce chaotic oscillation). 
The uncoupled oscillator has two symmetric saddle foci ($\pm \sqrt{ab}, \pm{\sqrt{\frac{a}{b}}}, 0$) and a Lorenz system-like butterfly attractor
but no equilibrium origin. Two such identical Sprott oscillators are mutually coupled via $x_1-, y_1$-variables while a repulsive link is fed back to the first oscillator via $x_3$-variable. The coupled system is,
%\begin{subequations}
%\begin{eqnarray}
$\dot{x_1}=x_2x_3+\epsilon(y_1-x_1)$,
$\dot{x_2}=x_1-bx_2 $,
$\dot{x_3}=a-x_1 x_2-\epsilon(x_3+y_3)$,
$\dot{y_1}=y_2y_3+\epsilon(x_1-y_1)$
$\dot{y_2}=y_1-by_2$,
$\dot{y_3}=a-y_1 y_2$,
%\end{eqnarray}
%\end{subequations}
where $p_1=p_2=1$ and $q_1=-1$ and $q_2=0$,
\begin{eqnarray}
D=
\begin{pmatrix}
 \epsilon_1 & 0 & 0  \\
0 & 0 & 0\\
0 & 0 & 0
  \end{pmatrix}
% \end{eqnarray}
%\begin{eqnarray}
;
Q=
\begin{pmatrix}
 0 & 0 &0  \\
0 & 0  &0\nonumber\\
0 & 0 & \epsilon_2
  \end{pmatrix}
 \end{eqnarray}

The equilibrium points of the coupled systems are, 
%[$x_1^*=\frac{(\epsilon^2-1)\pm\sqrt{(\epsilon^2-1)^2-4\epsilon^2}}{2}$; $x_2^*=x_1^*$; $x_3^*=\frac{\epsilon(x_1^*-1)}{x_1%^*}$; $y_1^*=1$ $y_2^*=1$; $y_3^*=\epsilon(1-x_1^*)$; ]\\
%[$x_1^*-\frac{-(\epsilon^2-1)\pm\sqrt{(\epsilon^2-1)^2-4\epsilon^2}}{2}$;  $x_2^*=x_1^*$; $x_3^*=\frac{\epsilon(x_1^*+1)}{x_1^*}$; $y_1^*=-1$ $y_2^*=-1$; $y_3^*=\epsilon(1+x_1^*)$; ]\\
 $x_1^*=b x_2^*$; $x_2=\pm {\frac{{\frac{a}{b}}-\epsilon_1 \epsilon_2 \pm\sqrt{(\frac{a}{b}-\epsilon_1 \epsilon_2)^2-4\epsilon_1 \epsilon_2\frac{a}{b}}}{2 \sqrt{\frac{a}{b}}}}$; $x_3^*=\frac{\epsilon_1(x_2^*-y_2^*)}{x_2^*}$; $y_1^*=\pm \sqrt{ab}$; $y_2^*=\pm\sqrt{\frac{a}{b}}$; $y_3^*=\frac{\epsilon_1(y_2^*-x_2^*)}{y_2^*}$.
\begin{figure}[h]
\subfigure(a)\includegraphics[height=6.5cm,width=8cm]{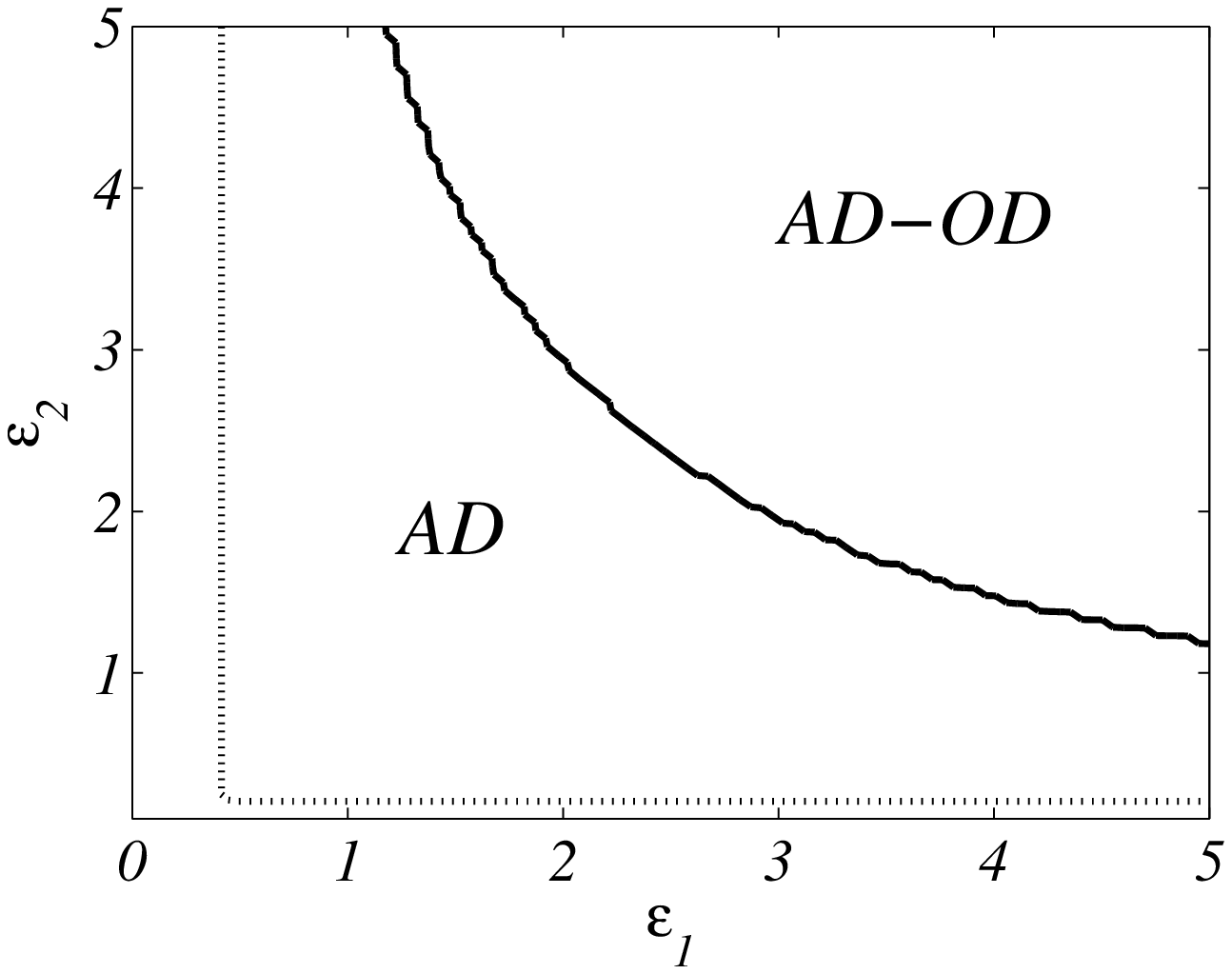}
\subfigure(b)\includegraphics[height=7cm,width=8.5cm]{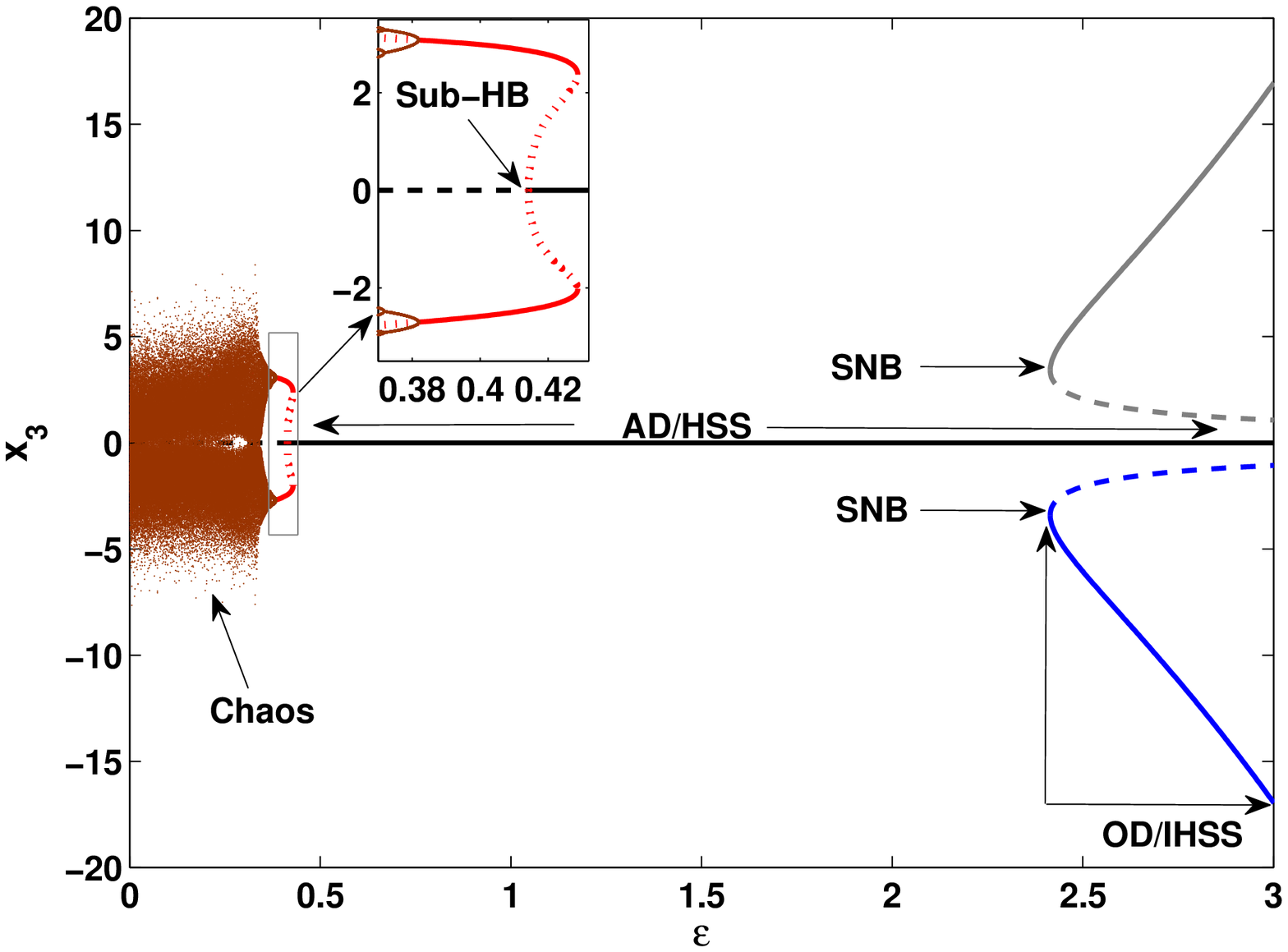}
\caption{Two diffusively coupled Sprott oscillators with one negative link ($a$=1, $b$=1). Phase diagram in $\epsilon_1$-$\epsilon_2$ space in (a) oscillatory region exists in the yellow region, AD emerges in the light red region and AD/OD in the dark brown region with clear transition boundaries. Bifurcation diagram using extrema of $x_3$ variable shows in (b) reverse-period doubling and subcritical HB (boxed region zoomed in upper inset). Symmetric blue and gray lines show origin of OD at a critical coupling with coexisting unstable steady states in dotted lines.  AD coexists with OD here. This is a particular case ($\epsilon_1=\epsilon_2$) along the diagonal line in Fig. 2(a) } \label{fig:sprott_bi}
\end{figure}
From numerical simulation of the eigenvalues of the {\it Jacobian} of the coupled system, AD and OD regimes are identified and plotted in the phase diagram Fig. {\ref{fig:sprott_bi}}(a) by varying $\epsilon_1$ and $\epsilon_2$. AD is identified by the zero-crossing of the real part of the complex conjugate eigenvalues while OD is recognized by the zero-crossing of the largest real eigenvalue.
A particular case of $\epsilon_1$ =$\epsilon_2$= $\epsilon$ is shown in a bifurcation diagram in Fig. {\ref {fig:sprott_bi}}(b) (using MatCont software package), where we plot extrema of the $x_3$ variable as a function of the coupling strength $\epsilon$. This case matches with the parameter set along the diagonal of the phase diagram. For lower values of $\epsilon$, the system shows chaotic oscillation (brown dots). With increase of $\epsilon$, the system shows reverse period-doubling (solid brown line in the inset) which merges with a limit cycle (solid red line) originated via subcritical HB at $\epsilon = 0.4142$.  An unstable fixed point (dashed black curve) always co-exists in this region which stabilizes at the subcritical HB point when the HSS origniates (solid black curve). The subcritical HB, of course, shows a hysteresis in the coupling range $\epsilon$=0.4142-0.4287. The HSS state orginating via subcritical HB continues to exist for higher values of $\epsilon$ until a transition to OD occurs at $\epsilon > 2.414$ via saddle-node bifurcation (SNB). SNB is indicated by two solid lines (solid blue and gray lines) coexisting with dotted (blue and grey) lines. The dotted lines are the coexisting unstable steady states indicating the origin of SNB at $\epsilon = 2.414$. The HSS coexists (zero black line) with the IHSS here in contrast to a previous report \cite{stefanovoska} where an unstable origin bifurcates into IHSS via SNB but the origin remains unstable. No transition from HSS to IHSS was observed although a negative mean-field coupling or a repulsive link was used there, possibly due to a symmetric use of the repulsive coupling.
Note that this example shows a general case of HSS where no equilibrium origin exists. The equilibrium points of the coupled system are [1,1,0, 1,1,0] and [-1,-1,0,-1,-1,0] for the selected parameters where the coupled system becomes stable at either of them and this can be clearly revealed if we plot the other state variables but not shown here. The $x_3$ variable always remain zero at stable nonzero equilibrium.

\begin{figure*}%[h]
%\vspace{-2.5cm}
%sprott_3osc_bif_node
%\subfigure(a)\label{fig:sprott_3osc_node}\includegraphics[height=5cm,width=4.cm]{./figures/node.eps}
%Sprott_3osc_superimage1

\subfigure(a)\includegraphics[height=6cm,width=7cm]{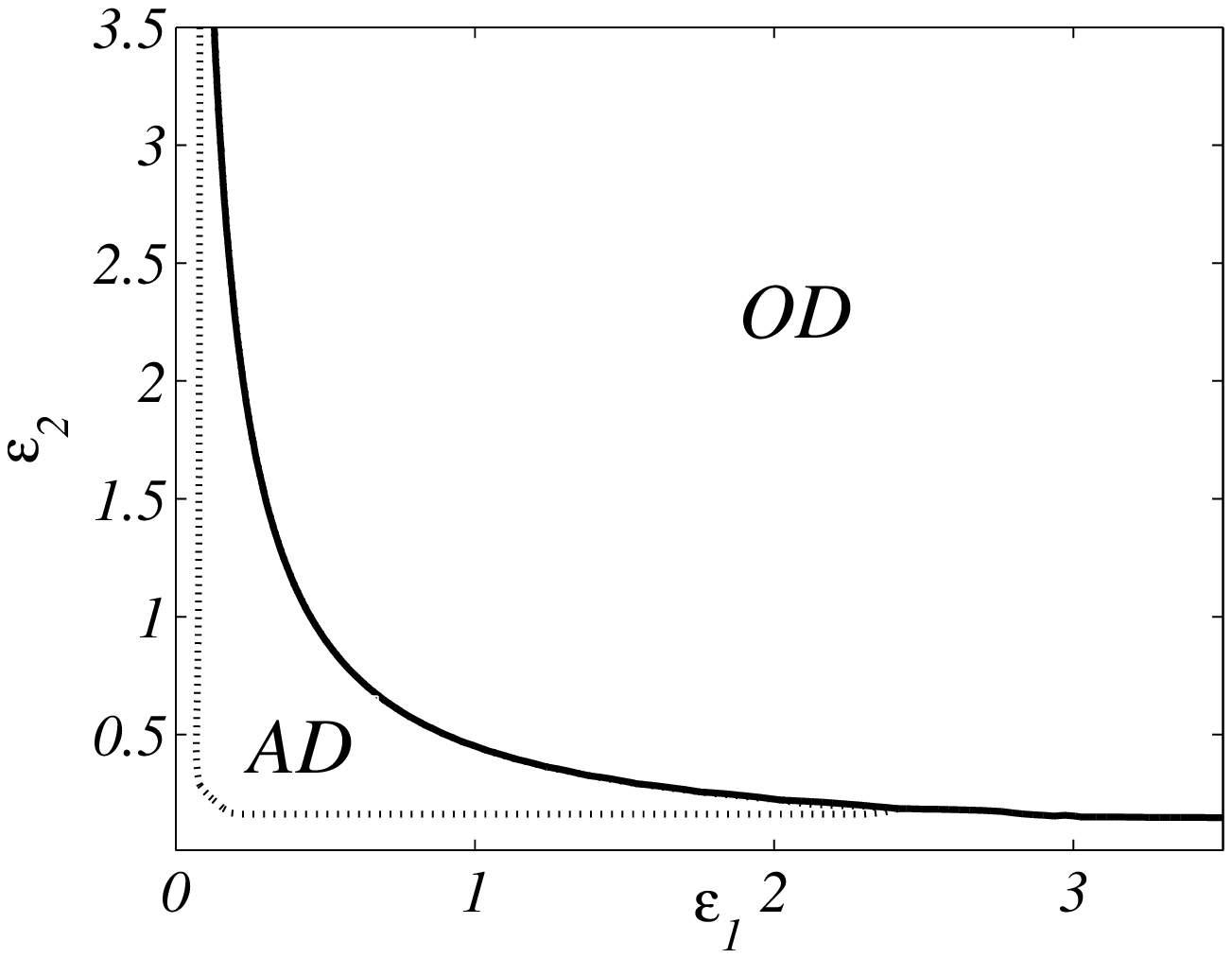}
\subfigure(b)\includegraphics[height=6cm,width=8.5cm]{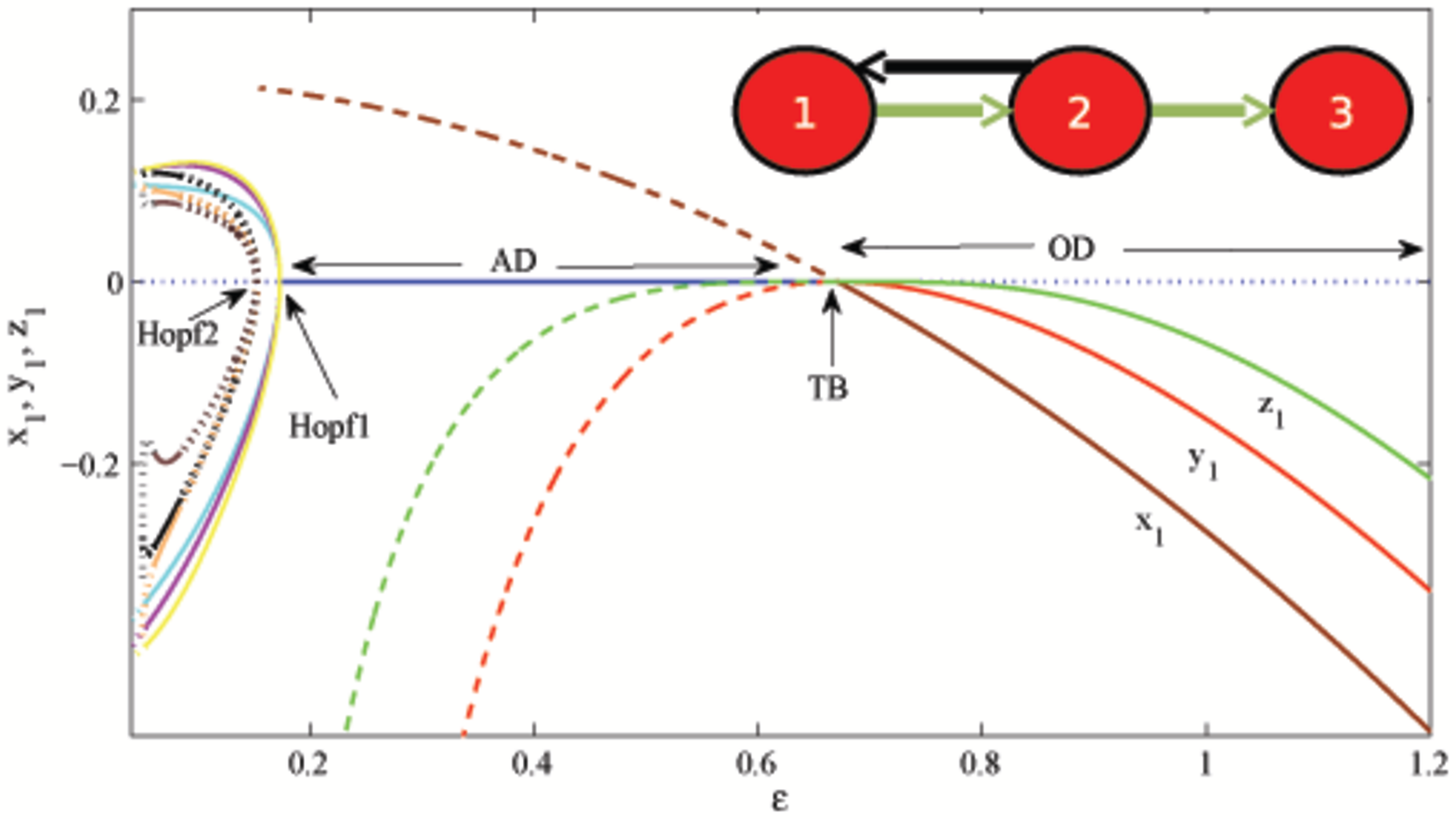}
\caption{ Three unidirectionally coupled Sprott oscillators with one repulsive link. (a) AD and OD has been shown as a function of $\epsilon_1$ and $\epsilon_2$: oscillatory region in yellow color, AD and AD/OD regions are demarcated by orange and brown regions,(b) extrema of $x_1$, $y_1$ and $z_1$ plotted as a function of the coupling strength $\epsilon$ showing a transition from AD to OD via TB. The coupling scheme is shown at top right corner.} 
\label{fig:sprott_3osc_node}%\label{fig:sprott_3osc}
\end{figure*}

 Finally we consider a chain of three identical oscillators, one driving the other successively and one repulsive link returns to the first oscillator as shown at top right corner of Fig. {\ref{fig:sprott_3osc_node}}(b). The red circles represent the Sprott oscillators with green arrows for unidirectional diffusive coupling and black arrow for the repulsive link. This is the same system used earlier \cite{hens}, in addition, another driven Sprott system is included to form a chain. The uncoupled Sprott oscillator has a R\"{o}ssler-like coherent chaotic attractor for $a=0.225$.
%We will use the variables of the i th Sprott oscillator as: $[{\bf{X_i}}=x_1^i,x_2^i,x_3^i:]$.
The chain of Sprott oscillators with a repulsive link is, %({\ref{fig:sprott_3osc_node}}):
\begin{subequations}
\begin{eqnarray}
\dot{x_1}&=&-ax_2, \nonumber\\
\dot{x_2}&=&x_1+x_3-\epsilon_2(x_2+y_2),\nonumber \\
\dot{x_3}&=&x_1+x_2^2-x_3\nonumber\\
\dot{y_1}&=&-ay_2+\epsilon_1(x_1-y_1),\nonumber\\
\dot{y_2}&=&y_1+y_3,\nonumber\\
\dot{y_3}&=&y_1+y_2^2-y_3,\nonumber\\
\dot{z_1}&=&-az_2+\epsilon_1(y_1-z_1),\nonumber\\
\dot{z_2}&=&z_1+z_3,\nonumber\\
\dot{z_3}&=&z_1+z_2^2-z_3.\nonumber
\end{eqnarray}
\end{subequations}

 The coupled system has a trivial equilibrium origin, while other equilibrium points are,
$x_1^*=x_3^*$;$x_2^*=0$;$x_3=\frac{\epsilon_2 y_2^*}{2}$; $y_1^*=-y_3^*$; $y_2^*=\frac{2a-\epsilon_1 \epsilon_2}{\epsilon_2}$; $y_3^*=\frac{{y_2^*}^2}{2}$;
$z_1^*=-z_3^*$; $z_2^*=\frac{a}{\epsilon_1}\pm{\sqrt{(\frac{a}{\epsilon_1})^2-2{\frac{\epsilon_2}{\epsilon_1}} {y_1^*}}}$;
and $z_3^*=\frac{{z_2^*}^2}{2}$. 
The repulsive link creates an additional asymmetry in the unidirectional chain of Sprott oscillators that is necessary to induce a transition from AD to OD. We plot a phase diagram in the $\epsilon_1-\epsilon_2$ plane to show the AD (orange color) and OD (brown color) regimes separated by a black boundary indicating  transition line in Fig. {\ref{fig:sprott_3osc_node}}(a). The AD regime vanishes at a larger value of $\epsilon_1>2.4$ for a smaller value of $\epsilon_2$; only OD regime exists there. A typical case is presented in Fig. {\ref{fig:sprott_3osc_node}}(b) (for $\epsilon_1=\epsilon_2$) where HSS appears at a Hopf point; in fact there are two Hopf points (Hopf1 and Hopf2) at left but HSS (solid blue line) appears at Hopf1 point but coexists with three unstable equilibrium points (brown, green and red dotted lines). The chaotic dynamics becomes period-1 for a very weak coupling via reverse-period doubling. Finally, HSS transits to IHSS via TB at a larger coupling when the equilibrium origin becomes unstable (dotted grey line) and the unstable equilibrium points become stable as indicated by the solid lines (green, red and brown). Three similar state variables $x_1, y_1$ and $z_1$ of the three coupled oscillators are plotted to show the transition. %We mention that if we increase the number of oscillators in the chain upto 30 oscillators, one repulsive is still able to induce HSS and a transition to IHSS. We plan to explore this with further details in the future.
 \par In summary, we have investigated the transition from amplitude death to oscillation death in three different systems representing  two identical oscillators under diffusive coupling and with an additional repulsive coupling.  The objective was to gain a better and more general understanding of this transition mechanism and investigate its relation to the complexity of the dynamics of the individual oscillatory units. With this in mind we chose two of the above systems to be of the chaotic type unlike previous studies \cite {koseska2, zou} which had all been confined to limit-cycle type systems. 
Our results show that the transition can occur through diverse routes. 
%We searched for a general mechanism of such transition in different dynamical systems, limit cycle and chaotic, however, found three diverse routes. 
In the limit cycle Van Der Pol system, the AD to OD transition follows the pitchfork bifurcation as reported earlier \cite {hens} for a Landau-Stuart system. However, in chaotic systems, we found two additional routes, transcritical bifurcation in one chaotic Sprott system (which had a Rossler like attractor) and a saddle-node bifurcation in another chaotic Sprott system (that had a Lorenz like attractor). 
%However, all these bifurcation routes, the pitchfork, transcritical and the saddle-node bifurcation all belong to the broad class of Turing instability. 
To the best of our best knowledge, the existence of such diverse transition routes between amplitude to oscillation death have not been reported so far and provides a fundamentally new insight into this phenomenon.  Our approach in studying this transition has also been different from those used in recent reports \cite {koseska2, zou}. Instead of a parameter perturbation, we have used an additional repulsive feedback link to create an asymmetry or heterogeneity in a synchronized set of oscillators under diffusive coupling and have varied the coupling strength when the transition appears. Our model and its results can have interesting practical consequences. For example, the repulsive link, which may evolve in time in a synchronized network of dynamical systems, can model an inhibitory coupling in a cell signaling network \cite{kondor} or a local fault in a synchronized power-grid network \cite{motter}. Thus the emergence of such a repulsive link may give rise to a destructive effect in the form of a fault or a disease in a synchronized network.  The fault can also be considered as an overloading in a synchronized power-grid that can lead to a complete black-out as a manifestation of amplitude death. Alternatively, in the case of a spreading epidemic \cite{wu}, a local awareness campaign in a population network may be considered as a local repulsive link that can play a constructive role in curbing the spread of an epidemic. This is also similar to an amplitude death state. On the other hand, in a cell signaling network, where a robust ground state is desirable, the presence of a repulsive link representing a disease like cancer can split the robust ground state into undesirable multiple stable steady states which is analogous to a transition to oscillation death. Our discovery of the diverse AD-OD transition routes can prove helpful in providing a better representation of such phenomenon in synchronized dynamical networks and their application to real world issues.
%In perspectives of such real world issues, our future plan is to explore amplitude death and its possible transition to oscillation death as a destructive or a constructive effect in synchronized %dynamical networks.

C.R.H. is supported by the CSIR Network project GENESIS (project no. BSC0121). S.K.D. and P.K.R acknowledge individual support by CSIR Emeritus scientist schemes. The authors like to thank Saugata Bhattacharyya for very useful discussions.

\end{document}